# Compressing Complexity: A Critical Synthesis of Structural, Analytical, and Data-Driven Dimensionality Reduction in Dynamical Networks

Zebiao Li [a], XueYing Wu [b], Chengyi Tu [b, *]

[a] Keyi College of Zhejiang Sci-Tech University; Shaoxing, 312369, China

[b] School of Economics and Management, Zhejiang Sci-Tech University; Hangzhou, 310018, China

[*] Corresponding author. Email: chengyitu1986@gmail.com

**Abstract**

The contemporary scientific landscape is characterized by a "curse of dimensionality," where our capacity to collect high-dimensional network data frequently outstrips our ability to computationally simulate or intuitively comprehend the underlying dynamics. This review provides a comprehensive synthesis of the methodologies developed to resolve this paradox by extracting low-dimensional "macroscopic theories" from complex systems. We classify these approaches into three distinct methodological lineages: Structural Coarse-Graining, which utilizes spectral and topological renormalization to physically contract the network graph; Analytical-Based Reduction, which employs rigorous ansatzes (such as Watanabe-Strogatz and Ott-Antonsen) and moment closures to derive reduced differential equations ; and Data-Driven Reduction, which leverages manifold learning and operator-theoretic frameworks (e.g., Koopman analysis) to infer latent dynamics from observational trajectories. We posit that the selection of a reduction strategy is governed by a fundamental "No Free Lunch" theorem, establishing a Pareto frontier between computational tractability and physical fidelity. Furthermore, we identify a growing epistemological schism between equation-based derivations that preserve causal mechanisms and black-box inference that prioritizes prediction. We conclude by discussing emerging frontiers, specifically the necessity of Higher-Order Laplacian Renormalization for simplicial complexes and the development of hybrid "Scientific Machine Learning" architectures—such as Neural ODEs—that fuse analytical priors with deep learning to solve the closure problem.

# 1 Introduction

The contemporary scientific landscape is defined by a paradox of scale. In domains ranging from systems biology and neuroscience to social infrastructure and power grids, our capacity to collect high-dimensional data has vastly outstripped our ability to intuitively comprehend or computationally simulate the underlying dynamics. We are increasingly confronted with systems composed of $N \approx 10^6$ to $10^9$ interacting components, where the sheer magnitude of the state space renders direct analysis intractable[1,2]. This challenge, often termed the "curse of dimensionality," necessitates a fundamental shift in our modeling paradigm. The central scientific objective is no longer merely to catalogue the microscopic interactions of every node, but to extract a low-dimensional description—a "macroscopic theory"—that faithfully reproduces the system's collective behavior[3,4]. This review provides a critical synthesis of the methodologies developed to achieve this compression, exploring the intersection of graph theory, non-linear dynamics, and machine learning[1,2,5].

The problem of dimensionality reduction in complex networks is conceptually analogous to real-space renormalization in statistical physics. Just as statistical mechanics bridges the gap between particle kinetics and thermodynamic variables, network reduction seeks to decimate microscopic degrees of freedom to reveal effective macroscopic dynamics[3]. The central objective is to construct a reduced representation—whether a smaller graph $\tilde{G}$, a set of effective differential equations, or a latent manifold—that preserves specific invariants of the original system, such as spectral properties, synchronization thresholds, or epidemic spreading rates[1,3,6,7].

Historically, approaches to this problem have bifurcated into distinct methodological lineages, each prioritizing different aspects of the fidelity-scalability trade-off. The first lineage, Structural Coarse-Graining, addresses reduction directly at the level of network topology. This paradigm aims to map the original graph to a reduced graph $\tilde{G} = (\tilde{V}, \tilde{E})$ with significantly fewer supernodes[3,8]. The second lineage, Analytical-Based Reduction, seeks to compress the dynamical state space itself rather than the graph topology[7,9]. These methodologies derive low-dimensional differential equations that govern the macroscopic observables of the system. The third and most rapidly evolving lineage is Data-Driven Reduction. Recognizing that explicit governing equations are often unknown in empirical systems, this approach infers low-dimensional manifolds directly from observational data[4,10].

This review posits that the selection of a reduction methodology is governed by a "No Free Lunch" theorem: a Pareto frontier exists between computational tractability and physical fidelity. Exact symmetry-based reductions offer perfect fidelity but are brittle to structural noise; spectral methods offer rigorous guarantees but scale poorly; and data-driven methods offer massive scalability but often act as phenomenological "black boxes"[7,9,11]. Furthermore, we argue that the field is currently witnessing an epistemological schism between equation-based derivations, which preserve causal mechanisms, and data-driven inference, which prioritizes predictive accuracy[1,2,12].

The following sections provide a comprehensive taxonomy of these methodologies and discuss emerging frontiers, including the necessity of moving beyond pairwise interactions to Higher-Order Laplacian Renormalization for simplicial complexes, and the development of hybrid architectures that fuse analytical priors with deep learning to solve the closure problem[5,13]. Through this synthesis, we aim to provide a roadmap for the rigorous compression of complex systems, enabling the discovery of the effective laws that govern high-dimensional chaos[6,14,15].

# 2 Structural Coarse-Graining Reduction

Structural coarse graining addresses the problem of dimensionality reduction directly at the level of the network topology[8,16]. The central objective of this paradigm is to construct a reduced graph $\tilde{G}=(\tilde{V},\tilde{E})$ — where the number of supernodes $|\tilde{V}|$ is significantly smaller than the original node count $|V|$ —that faithfully preserves specific spectral, dynamical, or topological invariants of the original system[17] (see Figure 2.1). This process is conceptually analogous to real-space renormalization in statistical physics, where microscopic degrees of freedom are iteratively decimated or aggregated to reveal macroscopic effective dynamics[18-20].

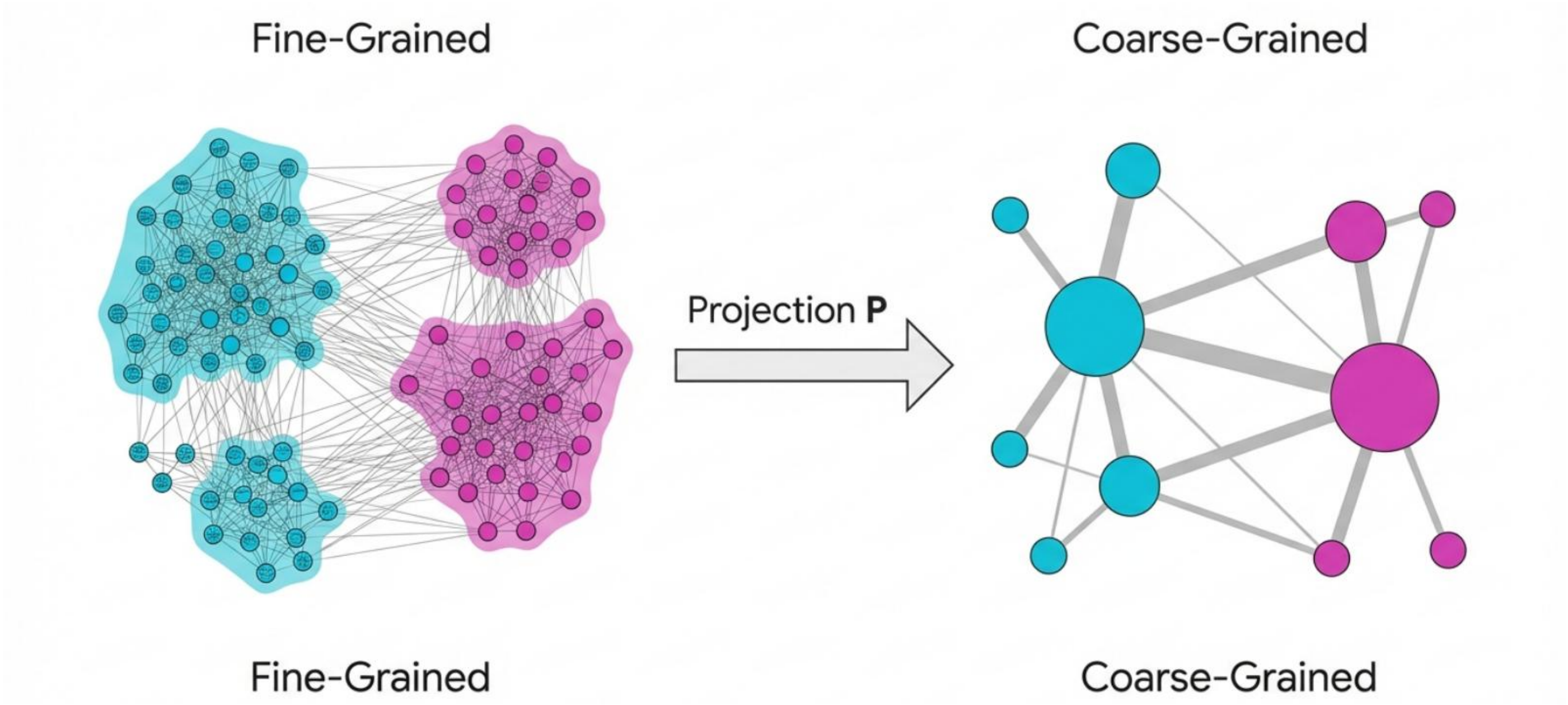

**Figure 2.1: Schematic illustration of the structural coarse-graining principle.** The original high-dimensional network (left) is partitioned into distinct clusters (colored regions) defined by structural or dynamical similarities. A projection operator maps these partitions to a reduced, low-dimensional network (right). In this renormalized topology, supernodes denote the aggregate behavior of the underlying clusters, while superedges represent the effective coupling between them.

## 2.1 The Gfeller-De Los Rios Framework

Among the diverse strategies employed for network dimensionality reduction, Spectral Coarse Graining (SCG) distinguishes itself as the most mathematically rigorous framework for the preservation of linear dynamics[8,16]. Developed principally by Gfeller and De Los Rios, SCG explicitly targets the conservation of the graph Laplacian's spectrum. By ensuring that the dominant eigenvalues and eigenvectors of the reduced network closely approximate those of the original, SCG guarantees that fundamental global properties—specifically diffusion rates, Mean First Passage Times (MFPT), and synchronization thresholds—remain invariant under the reduction[8,21].

The fundamental objective of SCG is the determination of a reduced network of size $\tilde{N} \ll N$ and an associated projection operator $\mathbf{P}$, such that the stochastic behavior of the original system is faithfully replicated in the lower-dimensional representation[16,22]. Conceptually, this mirrors the logic of the renormalization group in statistical physics, yet it relies on spectral similarity rather than spatial or topological proximity[19,20]. The methodological pipeline proceeds through following distinct stages (see Figure 2.2):

1. **Eigen-Decomposition[16,23]:** The procedure commences with the spectral decomposition of the relevant evolution operator (typically the random walk transition matrix $\mathbf{M}$ or the normalized Laplacian $\mathbf{L}$).

The first $k$ non-trivial eigenvectors are computed, defining the "slow modes" or the long-term relaxation dynamics of the system.

2. **Spectral Clustering[16,24]:** This step maps the topological space to a geometric one. Each node $i$ is assigned a coordinate vector $\mathbf{v}_i \in \mathbb{R}^k$ derived from the components of the first $k$ eigenvectors. In this spectral embedding, nodes are positioned not by their geodesic distance, but by their dynamical similarity; nodes that play equivalent roles in diffusion processes will be clustered proximally. A standard clustering algorithm (e.g., $k$-means) partitions the spectral space into $\tilde{N}$ distinct clusters or "supernodes." This partition minimizes the variance within supernodes, ensuring that grouped nodes share nearly identical probability transition profiles.
3. **Construct Projectors[16,25]:** The partition defines a projection matrix $\mathbf{P}$ ($\tilde{N} \times N$) and a lifting (or prolongation) matrix $\mathbf{P}^+$ ($N \times \tilde{N}$). $\mathbf{P}$ aggregates the probability mass of nodes within a cluster. $\mathbf{P}^+$ redistributes the mass from a supernode back to the individual nodes, typically weighted by the local stationary distribution $\pi$.
4. **Reduction Operator[16,25]:** The coarse-grained operator is derived via the similarity transformation $\mathbf{M} = \mathbf{PMP}^+$. Finally, $\mathbf{M}$ is interpreted as the weighted adjacency matrix of the reduced graph. Since $\mathbf{M}$ may be dense, a sparsification step is often required to render the reduced topology physically interpretable.

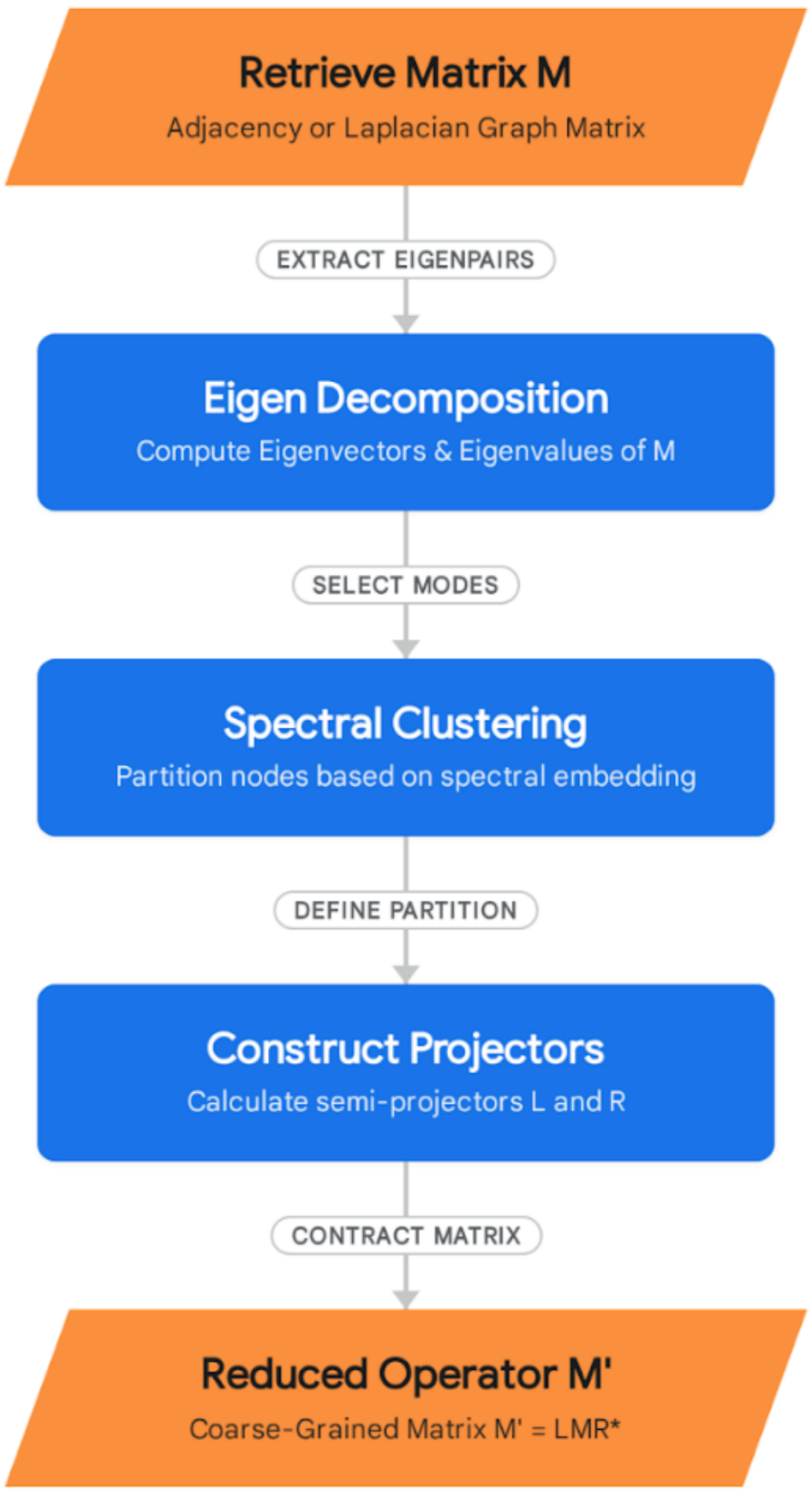

**Figure 2.2: Workflow of the Spectral Coarse Graining algorithm.** The schematic illustrates the computational pipeline, distinguishing between data input/output states (orange) and algorithmic transformations (blue). The process initiates with the retrieval of the network's representative matrix $M$ (Adjacency or Laplacian). Eigen Decomposition is performed to extract eigenpairs, identifying the slow modes that define the system's low-dimensional embedding. Nodes are subsequently partitioned within this spectral space to construct the semi-projectors ($L$ and $R$), which are used to contract the original matrix into the reduced coarse-grained operator $M'$.

The cardinal virtue of SCG is its isospectral fidelity—its ability to preserve the global relaxation timescales of the network. In the context of phase oscillator synchronization (e.g., the Kuramoto model), the linear stability of the synchronized state is governed by the eigenratio $\lambda_2 / \lambda_N$ of the Laplacian. SCG is designed to keep the algebraic connectivity ($\lambda_2$, the smallest non-zero eigenvalue) invariant[8,16,26]. Consequently, the critical coupling strength required for synchronization in the reduced model matches that of the original system with high precision.

Comparative analyses demonstrate that SCG significantly outperforms "naive" topological contractions (e.g., merging direct neighbors) or random groupings[21]. This is particularly evident in bipartite networks (e.g., client-server or metabolite-reaction graphs). Standard topological projections often obliterate the specific spectral signature required for accurate random walk analysis on bipartite structures. In contrast, modified SCG algorithms have been developed to explicitly handle these geometries, ensuring that the Mean First Passage Time (MFPT) of random walkers on the reduced graph deviates negligibly from the full simulation[21,25].

Despite its theoretical elegance, the practical application of SCG to massive, complex networks is constrained by computational and interpretative challenges.

- **Computational Complexity[16,23]:** The primary bottleneck lies in the eigen-decomposition. Computing the full spectrum is an $O(N^3)$ operation. While sparse iterative solvers (such as the Lanczos or Arnoldi algorithms) reduce this to approximately $O(k \cdot E)$—where $E$ is the number of edges—convergence is heavily dependent on the spectral gap. For networks with small spectral gaps (common in modular social networks), convergence slows dramatically. For graphs exceeding $10^7$ nodes, even this reduced complexity can be prohibitive.
- **Interpretation of Supernodes[23,27]:** The clusters generated by SCG are defined by mathematical similarity in eigenvector space, not necessarily by spatial locality. While these clusters often align with modular communities, they can occasionally group nodes that are geographically distant but functionally symmetric (e.g., peripheral nodes in a star graph). This non-locality complicates the physical interpretation of the reduced model, particularly in infrastructure engineering where spatial contiguity is paramount.
- **Matrix Realizability[19,20]:** The reduced operator $\mathbf{M}$ is mathematically optimal but not always physically realizable as a simple graph. The reduction process often generates a fully connected (dense) matrix of supernodes, including heavy self-loops. Furthermore, depending on the orthogonality of the projection, the resulting weights may be numerically negative, defying standard probabilistic interpretations (e.g., negative transition probabilities) and requiring ad-hoc truncation or regularization.

## 2.2 Exact Reductions: Symmetry and Automorphisms

While SCG accepts approximation errors to achieve significant reduction, a parallel branch of research focuses on exact dimensionality reduction. This approach relies on the rigorous mathematical properties of graph symmetries and automorphism groups[27].

A symmetry (or automorphism) of a network is a permutation of its nodes $\pi : V \to V$ that preserves the adjacency structure: $(u,v) \in E \Leftrightarrow (\pi(u), \pi(v)) \in E$. If such a symmetry exists mapping node $u$ to node $v$, the two nodes are structurally indistinguishable. The set of all such symmetries forms the Automorphism

Group of the graph, $\mathrm{Aut}(G)$. Nodes that can be mapped to one another by some element of the automorphism group form "orbits." The key insight of exact reduction is that nodes in the same orbit must, under deterministic dynamics with identical intrinsic parameters, evolve identically. They can therefore be collapsed into a single node in a quotient graph[23]. The dynamics on this quotient graph are an exact projection of the full system's dynamics; no information is lost regarding the collective behavior of the orbits.

The power of symmetry-based reduction lies in its potential for massive state-space compression without accuracy loss. Studies of real-world networks—including biological, social, and technological graphs—have revealed that "symmetry-induced lumping" can achieve compression ratios as high as $10^{12}$ in specific cases, such as animal social networks or highly regularized infrastructure grids[18,23]. This exactness is particularly valuable for stability analysis. Since the reduction is mathematically exact, any bifurcation (instability) detected in the reduced model corresponds to a genuine physical instability in the full system[23,28]. This provides a robust tool for analyzing the stability of synchronized states in large oscillator networks, where approximate methods might introduce spurious artifacts.

The fundamental trade-off in structural coarse graining is between the exactness of the reduction and the extent of the compression. Symmetry-based methods occupy one extreme of this spectrum: they offer zero error but are often limited in applicability. Real-world networks are rarely perfectly symmetric; they are plagued by noise, missing edges, and heterogeneity. A single misplaced edge can break the symmetry of a large cluster, destroying the automorphism group and rendering exact reduction impossible[23,27]. In contrast, SCG and similar spectral methods are robust to disorder. They group nodes that are "approximately" symmetric (spectrally similar), accepting a small error in exchange for a guaranteed reduction to any desired size $n$[16,21].

## 2.3 Iterative Structural Coarse Graining

As a response to the prohibitively high computational demands of spectral methods ($O(N^3)$) and the fragility of exact symmetry-based reductions, the Iterative Structural Coarse Graining (ISCG) framework has emerged as a robust alternative[16,23,29]. While Spectral Coarse Graining targets the preservation of linear diffusive modes, ISCG is specifically optimized for modeling non-linear contagion dynamics (e.g., SIS/SIR processes) on large-scale, heterogeneous networks[29,30].

ISCG operates on a principle fundamentally distinct from spectral renormalization. Rather than extracting global eigenmodes, it exploits the modular architecture of complex networks by identifying local structural motifs—specifically, dense subgraphs such as $k$-cliques (fully connected subgraphs of size $k$)[29].

The method is grounded in the physical insight that dense cliques function as "reservoirs" or "super-spreaders" in contagion processes. Within a fully connected clique, the mean time to saturation (full infection) is significantly shorter than the time required for the pathogen to diffuse between distinct communities. This separation of timescales allows the internal dynamics of the clique to be adiabatically eliminated; the internal state is less relevant to the macroscopic outbreak than the clique's aggregate coupling to the rest of the topology[29,30]. The ISCG algorithm proceeds via a recursive contraction strategy[29]:

1. **Motif Identification:** The algorithm scans the network topology to detect specific target mesostructures (e.g., 4-cliques, dense bipartite cores, or $k$-plexes).
2. **Structural Contraction:** Identified motifs are collapsed into single "supernodes." The internal degrees of freedom are integrated out, effectively treating the clique as a single dynamical unit.
3. **Renormalization of Flux:** The edges connecting the newly formed supernode to the remainder of the network are re-weighted. This step is critical to ensure that the total transmission probability (flux) between the collapsed clique and its neighbors remains invariant, preserving the global epidemic threshold.
4. **Iteration:** The process is repeated recursively on the reduced graph until a target system size $\tilde{N}$ or a specified error tolerance is reached.

A decisive advantage of ISCG is its computational scalability. Unlike SCG, which necessitates global matrix operations that scale cubicly with system size, the identification of local cliques is a parallelizable operation that can be performed independently on graph sub-sections. Consequently, ISCG can be successfully applied to massive networks with tens of millions of nodes—a regime where spectral decomposition becomes computationally intractable due to memory constraints[16,29].

ISCG has demonstrated particular efficacy in epidemiological applications, such as Sentinel Surveillance and targeted immunization[29,31]. By identifying the "super-spreader" cliques—which manifest as high-degree supernodes in the coarse-grained representation—public health interventions can be strategically directed toward these reservoirs. The reduced model faithfully reproduces the critical epidemic threshold and the final outbreak size, yet it is computationally lightweight enough to facilitate Monte Carlo simulations of thousands of intervention scenarios.

To address the "noise" inherent in empirical social data—where communities are rarely perfectly connected cliques—the framework has been extended to accommodate relaxed motifs such as $k$-plexes[29,30]. A $k$-plex of size $n$ is a subgraph where every node is connected to at least $n-k$ other nodes within the subgraph. This relaxation allows ISCG to coarse-grain social networks that are dense but imperfect (e.g., due to missing data or loose social ties), significantly broadening the domain of applicability beyond the strict constraints of clique-based topology.

## 2.4 Higher-Order and Geometric Renormalization

Traditional network science has predominantly modeled complex systems as collections of pairwise interactions (dyads). However, the dyadic paradigm is increasingly recognized as an insufficient approximation for many biological, neural, and social systems where interactions are fundamentally multi-way (polyadic)[18,20]. A metabolic reaction requiring three substrates, a simultaneous firing pattern within a neuronal assembly, or a collaborative social group cannot be losslessly decomposed into a sum of pairwise links. This realization has catalyzed a paradigmatic shift toward Higher-Order Networks, utilizing mathematical structures such as Simplicial Complexes and Hypergraphs, and necessitating the development of renormalization techniques that transcend simple node aggregation.

In the framework of simplicial complexes, interactions are formalized as simplices of varying dimensions: a 0-simplex represents a node, a 1-simplex an edge, a 2-simplex a filled triangle (a 3-way interaction), and so forth. Coarse-graining in this regime requires a Renormalization Group (RG) approach that operates natively across this hierarchy of dimensions. Recent theoretical breakthroughs have introduced "Higher-Order Laplacian Renormalization." This framework generalizes the spectral logic of Spectral Coarse Graining (SCG)[16,20]. Just as the standard graph Laplacian ( $L_0$ ) governs diffusive processes on nodes, the Hodge Laplacians ( $L_k$ ) govern the diffusion and solenoidal flows on $k$ -dimensional simplices (e.g., flux across edges or circulation around faces)[20,25]. The renormalization scheme proposed by Petri et al. leverages the spectral properties of these higher-order operators to cluster simplices rather than just nodes. By identifying spectral symmetries in the Hodge Laplacian, the method groups $k$ -simplices exhibiting similar flow dynamics into "super-simplices" (e.g., aggregating triangles into "super-triangles")[20,27]. This higher-order spectral analysis permits the detection of scale-invariant topological features that remain invisible to standard vertex-based metrics. For instance, a network may display scale-free behavior in its node degree distribution yet exhibit entirely distinct scaling universality in its higher-order connectivity (e.g., the distribution of 2-simplex participation)[20,32]. Only a renormalization flow acting on the full simplicial complex can decouple and reveal these distinct topological universality classes.

Parallel to these topological advances, Geometric Renormalization reconceptualizes complex networks as discrete approximations of underlying continuous manifolds[18,32]. This perspective is central to the emerging field of Geometric Deep Learning and the analysis of connectomes. The Geometric Renormalization Group (GRG) posits that the self-similarity frequently observed in complex networks—quantified by fractal box-counting dimensions—is a manifestation of a hidden geometric embedding space, most often characterized by hyperbolic curvature[18]. In this framework, coarse-graining is physically interpreted as a "zooming out" operation or a radial rescaling within the latent hyperbolic space. The renormalization flow naturally generates the ubiquitous structural properties of real-world networks: Clustering arises from the metric locality in the angular coordinates. Small-worldness is a consequence of the exponential expansion of space in the radial direction[18,23].

## 2.5 Computational Complexity and Comparative Synthesis

The selection of an optimal dimensionality reduction framework is governed by a fundamental trade-off between computational tractability and physical fidelity[16,23,29]. As detailed in the preceding sections, methods that offer rigorous spectral guarantees (such as SCG) often scale poorly with system size, rendering them infeasible for massive social or biological networks. Conversely, motif-based or neural approaches offer near-linear scalability but often lack the theoretical guarantees of their spectral counterparts. Table 2.1 provides a comprehensive taxonomy of the primary coarse-graining methodologies discussed in this review. It juxtaposes the algorithmic basis of each reduction against its asymptotic complexity, explicitly delineating the domain-specific conservation laws (e.g., effective resistance vs. epidemic threshold) that dictate their applicability.

**Table 2.1: Comparative Analysis of Network Reduction Methodologies.**

| Method | Basis of Reduction | Computational Complexity | Preserved Invariant / Property | Primary Domain & Application |
|---|---|---|---|---|
| Spectral Coarse Graining[8,16,21] | Spectral Similarity: Clustering of eigenvector components (slow modes). | High: $O(N^3)$ (full spectrum) or $O(k \cdot E)$ (iterative). Convergence depends on spectral gap. | Algebraic Connectivity ($\lambda_2$); Synchronization stability; Diffusion timescales. | Complex Systems: Phase oscillator networks; Random walks on graphs. |
| Exact Symmetry | Automorphism | Variable: | Exact topological | Theoretical |
| Reduction[23,27,28] | Group: Decomposition into independent orbits/subgroups. | Symmetry detection is GI-complete; practically feasible for sparse, symmetric graphs. | structure; Full transient dynamics; Soliton solutions. | Physics: Idealized lattices; Josephson junction arrays. |
| Iterative Structural Coarse Graining[29,30] | Local Motifs: Recursive contraction of dense subgraphs ( $k$ -cliques, $k$ -plexes). | Low / Scalable: $O(N)$ to $O(N \log N)$ (Parallelizable local search). | Epidemic thresholds ( $R_0$ ); Contagion outbreak size; Metastable states. | Epidemiology: SIS/SIR dynamics on social networks; Sentinel surveillance. |
| Higher-Order Laplacian Renormalization[18,20,32] | Hodge Theory: Spectral analysis of simplicial Laplacians ( $L_k$ ). | High: Scaling depends on the density of higher-order simplices (triangles, tetrahedra). | Topological scaling laws; Simplicial flow (flux/circulation); Betti numbers. | Network Neuroscience: Connectome topology; Scientific collaboration structures. |
| Neural Coarse Graining[33-35] | Differentiable Learning: Learned pooling functions (e.g., DiffPool, MinCutPool). | Hybrid: Training: High (GPU intensive); Inference: $O(N)$ (Inductive application). | Targeted non-linear dynamics; Thermodynamic consistency; Generative likelihood. | Machine Learning: Molecular dynamics; Drug discovery; Generative graph modeling. |

# 3 Analytical-Based Reduction

In contrast to structural coarse-graining, which achieves reduction by physically contracting the network topology, analytical-based reduction seeks to compress the dynamical state space itself. These methodologies operate not by clustering nodes, but by deriving a lower-dimensional set of differential equations that govern the macroscopic observables of the system[7,9,36]. The central premise is that the asymptotic behavior of high-dimensional coupled systems—such as the synchronization of $10^6$ neurons or the propagation of an epidemic—often collapses onto a low-dimensional invariant manifold[37-43].

## 3.1 Exact Dimensionality Reduction in Phase Oscillator Networks

The theoretical study of synchronization within large populations of coupled oscillators has historically provided the primary testbed for validating dimensionality reduction methodologies. The canonical mathematical object in this domain is the Kuramoto model, which describes the phase evolution of $N$ limit-cycle oscillators interacting via a sinusoidal coupling function. The seminal discovery that such systems—nominally possessing infinite degrees of freedom in the thermodynamic limit—could be exactly described by a finite, low-dimensional set of Ordinary Differential Equations (ODEs) marked a paradigm shift in non-linear dynamics[44]. This section details the two fundamental analytical reductions that make this possible.

In a landmark 1993 analysis, Watanabe and Strogatz (WS) uncovered a remarkable integrability in ensembles of identical phase oscillators subject to global coupling[9]. They investigated a general class of systems governed by the dynamical equation:

$$\dot{\theta}_i(t) = \omega(t) + \mathrm{Im}[H(t)e^{-i\theta_i}]$$

where $\omega(t)$ represents a common time-dependent natural frequency, and $H(t)$ denotes a common complex forcing function acting uniformly on all oscillators.

The profound insight of Watanabe and Strogatz was the identification of a fundamental symmetry: the equations of motion for the individual phases $\theta_i$ are invariant under the action of the Möbius group—the group of fractional linear transformations that map the unit disk onto itself. This symmetry implies that the $N$-dimensional phase space is foliated by 3-dimensional invariant submanifolds (alongside $N-3$ constants of motion)[9,45]. The WS transformation explicitly maps the $N$ microscopic phases $\theta_i$ to three macroscopic variables—amplitude $\rho$, phase $\Phi$, and angle $\Psi$—while the remaining degrees of freedom are frozen as constants $\psi_i$. Consequently, the evolution of the entire ensemble, regardless of size ($N \geq 3$), is rigorously determined by the trajectory of the Möbius parameters themselves.

The WS transformation is "exact" in the strictest mathematical sense. Unlike asymptotic methods, it captures the full transient dynamics, enabling the description of non-equilibrium phenomena such as phase space "solitons" and chaotic waves in non-integrable limits. However, the applicability of the WS ansatz is constrained by its requirement for symmetry. The introduction of heterogeneous natural frequencies—a hallmark of real-world complexity—breaks the strict Möbius invariance, rendering the WS reduction an approximation rather than an exact solution in generic contexts[7].

In 2008, Ott and Antonsen (OA) introduced a reduction framework that, while restricting the admissible class of initial conditions, successfully solved the problem of heterogeneity. Their approach is tailored for the Kuramoto model with Lorentzian (Cauchy) distributed natural frequencies[36,46].

Consider the thermodynamic limit ( $N \to \infty$ ) where the system state is described by a probability density function $f(\omega,\theta,t)$. The evolution of this density is governed by the continuity equation:

$$\frac{\partial f}{\partial t} + \frac{\partial}{\partial \theta}(vf) = 0$$

where $v(\omega,\theta,t)$ is the velocity field determined by the coupling. Ott and Antonsen proposed a specific ansatz for the Fourier series representation of $f$. They posited that the Fourier coefficients $f_n(\omega,t)$ decay geometrically as powers of a single complex function $\alpha(\omega,t)$:

$$f(\omega,\theta,t) = \frac{g(\omega)}{2\pi}\left[1 + \sum_{n=1}^{\infty}(\alpha(\omega,t)^n e^{in\theta} + c.c.)\right]$$

subject to $|\alpha(\omega,t)|<1$. This ansatz effectively restricts the dynamics to a specific, invariant sub-manifold within the infinite-dimensional space of probability distributions.

The power of the OA ansatz is fully realized when coupled with a Lorentzian frequency distribution, $g(\omega) = \frac{\Delta}{\pi}$. By utilizing the residue theorem and analytically continuing $\alpha(\omega,t)$ into the complex lower half-plane, the integral defining the global order parameter $Z$ can be evaluated exactly at the pole $\omega = \omega_0 - i\Delta$. This procedure collapses the infinite hierarchy of moment equations into a single, closed, low-dimensional ODE for the Kuramoto order parameter $z(t)$:

$$\dot{z} = -(\Delta - i\omega_0)z + \frac{K}{2}(z - z^* z^2)$$

This 2D dynamical system captures the critical phase transitions, the synchronization threshold $K_c = 2\Delta$, and the bifurcation structure of the full infinite-dimensional system. Furthermore, stability analyses by Pikovsky and Rosenblum have demonstrated that the OA manifold is weakly attracting for Lorentzian distributions;

trajectories starting off the manifold asymptotically approach it, justifying the use of these reduced equations for analyzing long-term attractors even from generic initial conditions[47].

Real-world oscillatory networks, particularly power grids and certain neuronal models, often exhibit inertial effects, necessitating second-order differential equations. The Kuramoto model with inertia is formally defined as:

$$m\ddot{\theta}_i + D\dot{\theta}_i = \Omega_i + \frac{K}{N}\sum_{j=1}^{N}\sin(\theta_j - \theta_i)$$

where $m$ denotes inertia/mass and $D$ the damping coefficient.

Reducing this system presents a significant challenge as the phase space doubles in dimension (encompassing both positions $\theta_i$ and velocities $\dot{\theta}_i$). The standard OA ansatz, defined strictly for phase distributions, is insufficient. However, recent theoretical extensions have successfully generalized the reduction to the tangent bundle by assuming the joint distribution of phase and velocity follows a specific marginal form compatible with the OA constraints[48,49]. The resulting reduced model is no longer a single complex ODE, but a set of coupled ODEs governing the order parameter $z$ and the mean ensemble velocity. Crucially, this reduction preserves the rich dynamical features unique to inertial systems, such as bistability (the coexistence of synchronized and incoherent states) and hysteretic transitions. These reduced models have become indispensable for analyzing the "Swing Equation" in power grid dynamics, allowing for the analytical derivation of stability boundaries without the computational cost of simulating the full network[50,51].

## 3.2 Effective Dynamics in Heterogeneous Networks

While the exact reductions of Watanabe-Strogatz and Ott-Antonsen provide profound insights into synchronization manifolds, their rigorous application is frequently constrained to systems with all-to-all coupling or high degrees of symmetry. In contrast, the vast majority of real-world complex systems—ranging from biochemical metabolic pathways to the topology of the Internet—are characterized by sparse connectivity, heavy-tailed degree distributions, and modular architectures. To bridge this gap, Gao, Barzel, and Barabási (2016) introduced a universal framework designed to map multi-dimensional network dynamics onto a single, effective, one-dimensional equation[6].

The Gao-Barzel-Barabási (GBB) framework operates on the premise that, for a broad class of dynamical systems, the localized microscopic degrees of freedom can be compressed into a global "effective" mean field. Consider a canonical dynamical system evolving on a network of $N$ nodes:

$$\frac{dx_i}{dt} = F(x_i) + \sum_{j=1}^{N} A_{ij} G(x_i, x_j)$$

where $F(x_i)$ governs the self-dynamics (e.g., intrinsic growth, decay, or oscillation), while $G(x_i, x_j)$ defines the interaction kernel (e.g., diffusion, regulatory inhibition), weighted by the adjacency matrix elements $A_{ij}$.

The reduction methodology maps these $N$ coupled differential equations to a single state variable $x_{\text{eff}} = \mathcal{L}(\mathbf{x})$ representing the weighted mean activity of the system and a corresponding topological control parameter $\beta_{\text{eff}} = \mathcal{L}(\mathbf{s}^{in})$, where $\mathcal{L}(\mathbf{x}) = \frac{\langle \mathbf{s}^{out}, \mathbf{x} \rangle}{\langle \mathbf{s}^{out} \rangle}$ is a topology-weighted mean-field operator and $\mathbf{s}^{out}$ is the out-degree of network. This effective parameter is a scalar functional of the network's topology, encapsulating the weighted density and heterogeneity of the connections. In the case of unweighted networks with degree distribution $P(k)$, $\beta_{\text{eff}}$ is governed by the ratio of the first two moments of the degree distribution: $\beta_{\text{eff}} = \frac{\langle k^2 \rangle}{\langle k \rangle}$. In more general contexts involving weighted edges, this parameter is derived from the spectral properties of the adjacency matrix[52]. The resulting "Effective Equation" takes the closed form:

$$\frac{dx_{\text{eff}}}{dt} = F(x_{\text{eff}}) + \beta_{\text{eff}} G(x_{\text{eff}}, x_{\text{eff}})$$

This 1D ODE serves as a proxy for the global state of the high-dimensional network. The fixed points of this equation ($\dot{x}_{\text{eff}} = 0$) correspond to the macroscopic steady states of the complex system. By performing a bifurcation analysis on this reduced equation with respect to $\beta_{\text{eff}}$, one can analytically predict the resilience and tipping points of the entire network[53,54].

The most significant theoretical implication of the GBB framework is the emergence of universality. The reduction posits that the macroscopic resilience of a system is dictated primarily by the effective parameter $\beta_{\text{eff}}$, rather than the microscopic intricacies of the network topology. This hypothesis leads to a powerful prediction: two topologically distinct networks—for instance, a highly heterogeneous scale-free graph and a homogeneous Erdős-Rényi graph—will exhibit identical collapse trajectories if they map to the same $\beta_{\text{eff}}$ [6,40]: (1) Data Collapse. When the system state is plotted against raw control parameters (e.g., average degree $\langle k \rangle$), the bifurcation diagrams for different network topologies appear distinct and uncorrelated. (2) Universal Function. However, when the state is re-plotted against $\beta_{\text{eff}}$, these distinct curves collapse onto a single

"universal resilience function".

Recent theoretical advancements have significantly extended the applicability of the GBB framework. Notably, the identification of a novel boundary condition has established the effective limits of the framework while simultaneously offering quantitative metrics for predicting approximation fidelity[55]. Furthermore, the framework has been generalized to accommodate non-homogeneous dynamical mechanisms, specifically in scenarios where the functional forms of self-dynamics and coupling-dynamics exhibit heterogeneity across nodes[56]. Consequently, this reduction methodology has been successfully adapted to encompass discrete-time maps[57], stochastic differential equations[58], and systems governed by separable coupling dynamics[59], thereby broadening the scope of universality to a much wider class of complex systems.

While the GBB framework offers a powerful analytical tool, it remains a mean-field approximation. Its validity hinges on specific spectral assumptions: principally, that the system state is dominated by the leading eigenvector of the adjacency matrix and that the spectral gap is sufficient to dampen transverse modes. The reduction performs exceptionally well for dense networks or those exhibiting small spectral gaps. However, accuracy degrades in networks characterized by strong modularity, large spectral gaps, or localization phenomena[43,60]. In these regimes, specific "hubs" or local clusters may deviate significantly from the global mean field. Under such conditions, the system does not undergo a uniform transition; rather, it may exhibit partial collapse where specific modules fail while others persist—a complex, fragmented state that the single variable $x_{\mathrm{eff}}$ cannot capture[61,62].

To rectify these limitations, recent hybrid methodologies such as G-LED (Generative Learning of Effective Dynamics) have been proposed. G-LED integrates the structural priors of the GBB framework with modern machine learning (specifically auto-regressive attention models)[63,64]. Instead of relying solely on the analytical derivation of $\beta_{\mathrm{eff}}$ —which may suffer from truncation errors—G-LED "learns" the effective dynamics directly from simulation data. By correcting for the non-linearities and higher-order correlations that the analytical mean-field ignores, this approach represents a synthesis of rigorous equation-based modeling with the adaptive accuracy of data-driven inference[65].

## 3.3 Mean-Field and Moment Closures

When the rigorous symmetries required for exact reductions (as seen in the Watanabe-Strogatz or Ott-Antonsen formalisms) are absent, and the system does not admit a clear geometric embedding, researchers resort to statistical approximations derived from the Master Equation (see Figure 3.1). These methods are particularly indispensable for discrete-state dynamics—such as epidemic spreading (SIS, SIR models) or Ising spin systems—where the state space is combinatorial ( $2^N$ ) rather than continuous[66-68].

The foundational approximation in this domain is Heterogeneous Mean-Field (HMF) Theory. HMF operates on a coarse-graining assumption: that all nodes possessing the same degree $k$ are statistically indistinguishable. Consequently, the dimensionality of the system is reduced from $N$ individual node equations to $K_{\max}$ class-level equations, where the state variable $\rho_k(t)$ represents the density of active (e.g., infected) nodes within degree class $k$.

For a standard Susceptible-Infected-Susceptible (SIS) epidemic model, the exact Markovian dynamics are approximated by the coupled differential equations:

$$\frac{d\rho_k}{dt} = -\mu\rho_k + \lambda k(1-\rho_k)\Theta(\rho)$$

where $\mu$ is the recovery rate, $\lambda$ is the infection probability per contact, and $\Theta(\rho)$ represents the global mean-field probability that a random link points to an infected neighbor[69-71]. This probability is calculated as a weighted average over all degree classes, $\Theta(\rho) = \frac{\sum kP(k)\rho_k}{\langle k \rangle}$.

HMF has been instrumental in uncovering fundamental topological effects, most notably the vanishing epidemic threshold in scale-free networks ( $\lambda_c \to 0$ as $N \to \infty$ for degree exponents $2 < \gamma \le 3$ ). However, HMF implicitly relies on the "annealed network" approximation: it treats the network as a dynamic object where edges are constantly rewired to preserve the degree distribution. By averaging out the static connectivity, HMF ignores local dynamical correlations and clustering. Consequently, it frequently yields quantitative errors when applied to networks with high clustering coefficients or fixed, sparse topologies[69,71].

To rectify the correlation-blindness of HMF, Pair Approximation (PA) extends the state space to track not just node states ([S], [I]), but edge states ([SS], [SI], [II]). This extension acknowledges that the state of a node is strongly correlated with the state of its immediate neighbors. However, this introduces the "moment closure problem"[72-74]: the evolution of single nodes depends on pairs; the evolution of pairs depends on triplets ([SSI], [ISI]); and so forth, generating an infinite hierarchy of moment equations:

$$\begin{aligned} \frac{d[S]}{dt} &= F([S],[SI]) \\ \frac{d[SI]}{dt} &= G([SI],[SSI],[ISI]) \\ &\cdots \end{aligned}$$

To render this system solvable, the hierarchy must be truncated by approximating higher-order terms as functions of lower-order terms. The canonical closure is the Kirkwood Approximation (or multiplicative closure), which assumes that, conditioned on the state of the central node, the states of its neighbors are independent[75,76]: $[ASI] \approx \frac{[AS][SI]}{[S]}$.

Advanced frameworks combine this moment expansion with degree heterogeneity[77,78]. Heterogeneous Pair Approximation (HPA) tracks edge densities stratified by the degrees of connected nodes, such as $\rho_{k,l}$ (the density of edges connecting a degree- $k$ node to a degree- $l$ node). By explicitly modeling the correlations between adjacent nodes, HPA significantly outperforms HMF in predicting phase transition points, particularly in modular or assortative networks where local interactions drive the global dynamics.

At the pinnacle of statistical approximation lies the Approximate Master Equation (AME) method. Rather than tracking only the first few moments (means and pairs), AME tracks the full probability distribution of the neighborhood states[37,42,79]. Specifically, it solves for the probability that a node of degree $k$ in state $S$ has exactly $m$ infected neighbors.

While computationally more demanding than standard Pair Approximation, AME provides near-exact agreement with stochastic Monte Carlo simulations for a vast array of dynamical processes on uncorrelated networks. It effectively solves the dimensionality reduction problem for binary-state dynamics by retaining the full distributional information of local neighborhoods, failing only when long-range loops or specific mesoscopic structures (like cliques) dominate the dynamics[43,79,80].

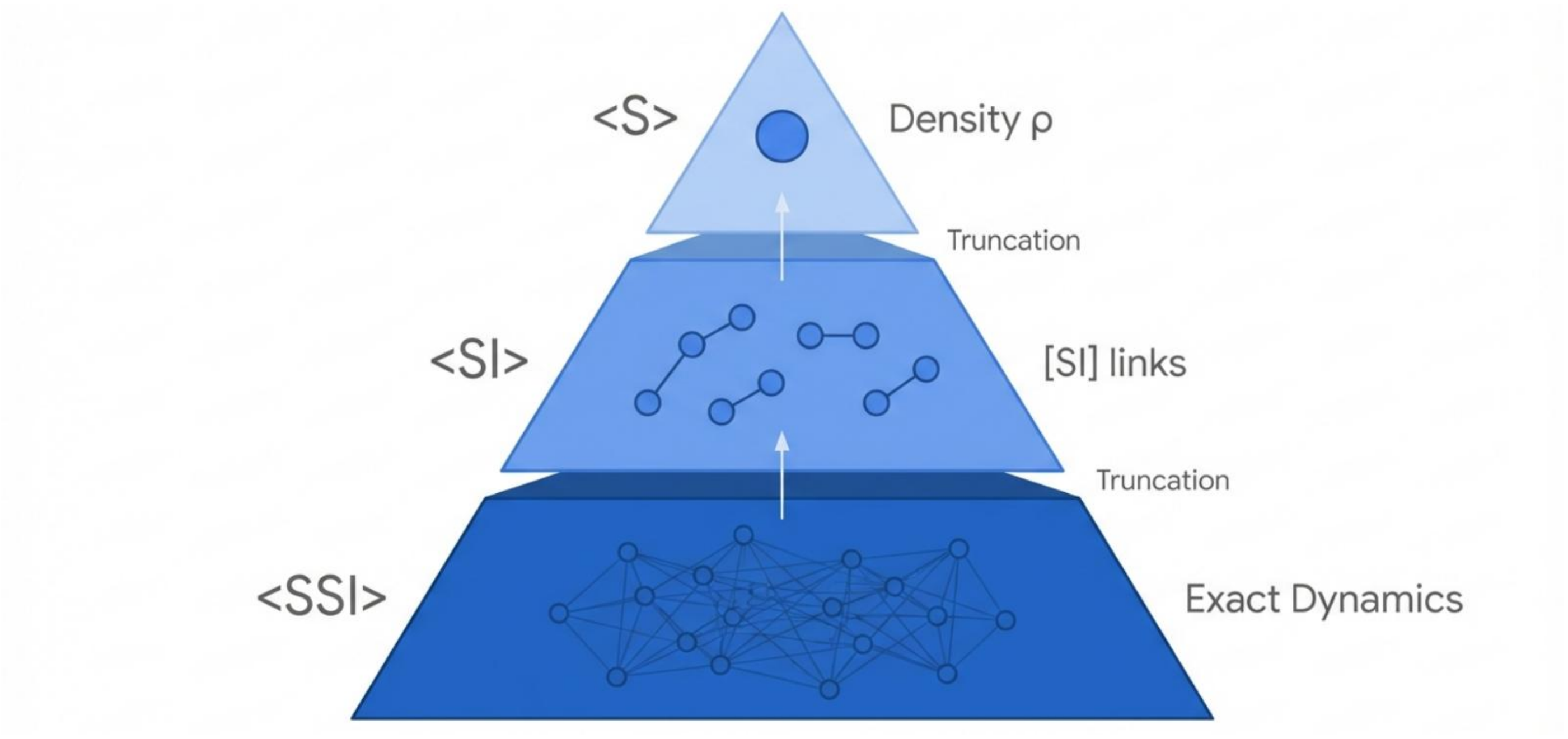

**Figure 3.1: Hierarchy of moment closure approximations.** The diagram depicts the systematic reduction

of the master equation state space. Starting from the exact dynamics at the base, higher levels introduce closure schemes to decouple the moment equations. The hierarchy transitions from triplet closure and pair approximation—which retain structural correlations in links ($\langle SI \rangle$)—to the mean-field approximation at the top, where the system is described solely by global density $\rho$.

## 3.4 Comparative Synthesis and Future Frontiers

The theoretical landscape of equation-based dimensionality reduction is characterized by a fundamental tension between universality—the capacity to describe diverse topologies and interaction rules—and fidelity—the mathematical rigor and precision of the reduced description[6].

This trade-off establishes a Pareto frontier of reduction methodologies[23,81]. On one end of the spectrum lie the symmetry-based reductions (Watanabe-Strogatz, Ott-Antonsen), which offer "exact" descriptions of infinite-dimensional systems. However, their reliance on specific algebraic structures (such as Möbius invariance) restricts their domain of validity to highly idealized networks[82,83]. On the opposite end lie the statistical and effective operator frameworks (Gao-Barzel-Barabási, HMF), which sacrifice analytical exactness in favor of broad applicability to the sparse, heterogeneous, and modular architectures encountered in real-world systems[6,84]. Table 3.1 provides a taxonomic summary of the discussed frameworks, delineating their theoretical bases, domains of applicability, and inherent limitations.

**Table 3.1 Comparative Analysis of Equation-Based Reduction Frameworks.**

| Framework | Mechanism | Applicability | Preserved Features | Limitations |
|---|---|---|---|---|
| Watanabe-Strogatz | Möbius Symmetry | Identical oscillators, Global coupling | Exact transient dynamics | Requires strict symmetry; fails with frequency heterogeneity |
| Ott-Antonsen | Invariant Manifold | Lorentzian frequencies, Sinusoidal coupling | Synchronization, Bifurcations | Strictly attracting only for Lorentzian; ignores some transients |
| Gao-Barzel-Barabási | Effective Operator | Weighted heterogeneous networks | Resilience points, Stability | Fails with strong localization, spectral gaps, or strong correlations |
| Heterogeneous | Degree-class | Scale-free networks | Epidemic | Ignores dynamical |
| Mean-Field | Averaging | (SIS/SIR) | Thresholds | correlations and local clustering |
| Pair Approximation | Moment Closure | Networks with clustering | Local Correlations | Combinatorial explosion of equations for higher accuracy |

# 4 Data-Driven Reduction

While the analytical methods discussed in Section 3 provide exact reductions for idealized networks, the complexity of real-world systems—characterized by stochasticity, structural heterogeneity, and often unknown governing equations—frequently renders such closed-form derivations intractable. Consequently, the field has increasingly pivoted toward data-driven dimensionality reduction[1,3]. These techniques infer low-dimensional representations directly from observational data, bypassing the requirement for explicit mechanistic models.

## 4.1 Spectral and Linear Reductions

The mathematical provenance of dimensionality reduction in network science is deeply rooted in spectral graph theory and linear algebra[85,86]. While the field has recently pivoted toward non-linear and neural-based manifold learning, these foundational spectral methods remain the canonical benchmarks for evaluation. They constitute the primary analytical apparatus for exploratory data analysis, offering rigorous, interpretable baselines against which the efficacy of modern, black-box techniques must be measured.

The stochastic behavior of diffusive processes on a network—ranging from heat dissipation and random walks to information cascades—is formally governed by the Graph Laplacian. This operator is defined as $\mathbf{L} = \mathbf{D} - \mathbf{A}$, where $\mathbf{D}$ denotes the degree matrix and $\mathbf{A}$ the adjacency matrix. The spectral decomposition of $\mathbf{L}$ yields a natural basis for dimensionality reduction, as the eigenvalues and eigenvectors encode the fundamental topology of the system. Specifically, the eigenvectors associated with the smallest non-zero eigenvalues (notably the Fiedler vector) delineate the slowest modes of relaxation within the system. These modes correspond to the "bottlenecks" of dynamic flow, effectively partitioning the network into metastable communities[8,16,22].

Laplacian Eigenmaps (LE) operationalize this spectral property. The algorithm derives a low-dimensional embedding by solving a generalized eigenvalue problem designed to preserve local neighborhood information [8]. Formally, LE minimizes the following objective functional:

$$\mathcal{O}_{LE} = \sum_{i,j} W_{ij} \| \mathbf{y}_i - \mathbf{y}_j \|^2$$

where $W_{ij}$ represents the affinity weight between nodes $i$ and $j$, and $\mathbf{y}$ denotes the coordinate vectors in the embedding space. This optimization ensures that nodes exhibiting strong dynamical coupling (high $W_{ij}$) are mapped to proximal coordinates in the low-dimensional manifold.

While LE excels at elucidating static community structures, it faces significant limitations in dynamical contexts[18,20,32]. Standard spectral methods rely on an implicit assumption of isometry—that the underlying manifold is globally or locally Euclidean. This assumption degrades rapidly in complex networks characterized by hyperbolic geometry (e.g., scale-free or hierarchical networks), where the volume of space expands exponentially rather than polynomially. Furthermore, LE focuses on the static projection of topological connectivity, often failing to capture the temporal evolution of a system's state trajectory.

Principal Component Analysis (PCA) serves as the ubiquitous baseline for data compression across scientific domains. Within the context of continuous dynamical systems and fluid mechanics, this method is frequently designated as Proper Orthogonal Decomposition (POD)[1,2]. The fundamental objective of PCA/POD is the identification of an orthogonal basis—the principal components—that maximizes the captured variance of the projected data. Mechanistically, given a centered data matrix $\mathbf{X}$, the method computes the covariance matrix $\mathbf{C} = \frac{1}{n}\mathbf{X}^T\mathbf{X}$. The principal directions are subsequently derived via the eigendecomposition $\mathbf{C} = \mathbf{V}\mathbf{\Lambda}\mathbf{V}^T$, or explicitly through Singular Value Decomposition (SVD), providing the optimal linear reconstruction of the state space.

Despite its computational tractability and analytic clarity, PCA exhibits critical shortcomings when applied to complex, non-linear network dynamics[1,87,88]: (1) The Linearity Constraint. PCA is strictly a linear transformation. If the system's attractor lies on a non-linear manifold (e.g., the canonical "Swiss roll" geometry), PCA creates a linear projection that cannot unfold the intrinsic structure. This results in a catastrophic overestimation of the intrinsic dimension and the spurious merging of distinct dynamical regimes that overlap only in the projection, not in phase space. (2) Insensitivity to Transient Dynamics. By design, PCA prioritizes high-energy (high-variance) modes. However, in many dynamical networks, critical precursors to phase transitions—such as bifurcation points or cascade initiations—manifest in low-variance modes (the "tail" of the eigenvalue spectrum). PCA effectively filters these subtle, information-rich signals as noise during truncation. (3) Reliance on Second-Order Statistics. The statistical justification for PCA assumes

Gaussian-distributed data. Real-world networks, however, frequently exhibit heavy-tailed degree distributions and intermittent activity patterns (burstiness). These phenomena violate the Gaussian assumption, rendering variance-maximization an insufficient proxy for information retention.

Multidimensional Scaling (MDS) extends the logic of dimensionality reduction by shifting the objective from variance maximization to distance preservation[1,89]. (1) Classical MDS (cMDS). When utilizing the Euclidean metric, cMDS is mathematically equivalent to PCA. It reconstructs embedding coordinates via the spectral decomposition of the Gram matrix derived from inter-point distances. (2) Non-Metric MDS. This variant relaxes the constraint of metric preservation in favor of topological consistency. It seeks to preserve the rank order of dissimilarities rather than their absolute magnitudes. This approach is particularly advantageous in social or biological networks where "distance" (e.g., genetic similarity or social influence) is an ordinal or qualitative construct rather than a rigorous metric space.

While MDS offers superior flexibility regarding distance metrics, it is hindered by severe computational bottlenecks[1,2]. The construction of the full pairwise distance matrix for a network of $N$ nodes necessitates $O(N^2)$ memory, while the subsequent eigendecomposition scales as $O(N^3)$. Perhaps more critically for dynamical systems, MDS lacks a direct out-of-sample extension (the Nyström extension problem). Projecting a new temporal state requires re-optimizing the entire embedding, rendering standard MDS impractical for the real-time monitoring of evolving networks.

## 4.2 Geometric Manifold Learning

To mitigate the limitations of linear spectral methods when applied to curved state spaces, the field has adopted a suite of geometric manifold learning algorithms. These methods rely on the Manifold Hypothesis: the assumption that while the high-dimensional data may be globally non-linear, it is locally isomorphic to a lower-dimensional Euclidean space (a tangent space). By effectively "stitching" together these local linear patches, these algorithms aim to recover the intrinsic non-linear degrees of freedom—conceptually "unrolling" the global manifold to reveal the latent dynamics[1,2,90].

A fundamental dichotomy in non-linear dimensionality reduction is the optimization trade-off between preserving global geometry (the overall topology and long-range dependencies of the trajectory) and local geometry (the nearest-neighbor relationships)[1,89].

Isomap (Isometric Mapping) extends the theoretical framework of MDS by redefining the metric of the ambient space. Recognizing that Euclidean distance often fails to respect the geometry of non-convex manifolds (e.g., cutting through the "void" of a Swiss roll), Isomap approximates the geodesic distance—the distance along the manifold's surface[90]. Mechanistically, the algorithm constructs a weighted neighborhood graph (typically a $k$-Nearest Neighbor graph) and computes the shortest path between all pairs of nodes using Dijkstra's or Floyd-Warshall's algorithm. The resulting geodesic distance matrix is then embedded via classical MDS. While Isomap is theoretically robust for recovering isometric embeddings, it is topologically brittle. It suffers acutely from "short-circuiting," where a single spurious edge—caused by noise—bridges two distant folds of the manifold. This topological violation collapses the geodesic distances, leading to a severe distortion of the global embedding[91].

In contrast, LLE abandons the attempt to estimate global geodesics, focusing exclusively on local preservation. The algorithm posits that every data point $\mathbf{x}_i$ and its neighbors lie on or close to a locally linear patch of the manifold. Consequently, $\mathbf{x}_i$ can be reconstructed as a linear combination of its $k$ nearest neighbors:

$$\mathbf{x}_i \approx \sum_j w_{ij} \mathbf{x}_j$$

LLE operates in two stages: first, minimizing the reconstruction error to solve for the weights $W_{ij}$; second, fixing these weights to solve for the low-dimensional coordinates $\mathbf{y}_i$ that best respect the local geometry. While LLE is immune to the short-circuiting that plagues Isomap, it introduces a different set of artifacts. By lacking a global constraint, LLE is prone to "covariance collapse" and aspect ratio distortion. For dynamical systems, this often manifests as the temporal compression of trajectories, where long periods of evolution are compressed into short segments in the embedding, obscuring rates of change[1,92].

Recent advances have shifted focus toward probabilistic and topological methods, which prioritize the separation of clusters and the visualization of high-dimensional structures.

t-Distributed Stochastic Neighbor Embedding (t-SNE) transforms the problem of embedding into one of probability distribution matching[93,94]. It converts high-dimensional Euclidean distances into conditional probabilities representing similarities, utilizing a Gaussian kernel in the high-dimensional space and a heavy-tailed Student's t-distribution in the low-dimensional space. The algorithm minimizes the Kullback-Leibler (KL) divergence between these two distributions. The heavy tail of the t-distribution effectively alleviates the "crowding problem," allowing t-SNE to reveal distinct dynamical regimes (e.g., separating "seizure" vs. "interictal" states in neural data) with exceptional clarity. t-SNE is inherently non-metric regarding global distances. It frequently "shatters" continuous time-series trajectories, fragmenting a smooth temporal evolution into disjoint clusters based on local density, thereby destroying the visual continuity essential for analyzing dynamical flows.

Uniform Manifold Approximation and Projection (UMAP) represents a significant theoretical leap, grounded in algebraic topology and Riemannian geometry[2] . It assumes the data is uniformly distributed on a locally connected Riemannian manifold. UMAP constructs a fuzzy simplicial set representation of the data and optimizes a cross-entropy loss to align the low-dimensional embedding with this topology. UMAP offers a superior balance between local and global structure compared to t-SNE and boasts deterministic scalability to millions of data points. However, standard UMAP remains a static algorithm; when applied to time series, it can produce the "spaghetti effect," where trajectories tangle or twist unnaturally due to the lack of temporal regularization[5,88].

To reconcile static manifold learning with dynamic data, recent innovations such as Directional Coherence Loss (DCL) and Temporal UMAP have been proposed[3,95,96]. These modifications explicitly incorporate the "arrow of time" into the objective function:

$$\mathcal{L}_{total} = \mathcal{L}_{topo} + \lambda \mathcal{L}_{temporal}$$

By penalizing embeddings that violate temporal continuity or directional smoothness, these methods ensure the low-dimensional vector field remains coherent, preserving the causal integrity of the network's evolution.

## 4.3 Operator-Theoretic Frameworks

Whereas geometric manifold learning targets the topological unfolding of the state space manifold, operator-theoretic frameworks propose a fundamental shift in perspective: moving from the evolution of states to the evolution of observables. This paradigm, grounded in the seminal work of Bernard Koopman (1931), posits that non-linear dynamical systems in a finite-dimensional state space can be equivalently represented as linear dynamical systems in an infinite-dimensional Hilbert space of observable functions[97]. This theoretical pillar has experienced a renaissance in the era of big data. It provides a rigorous bridge between the complexities of non-linear network dynamics and the well-established, powerful toolkit of linear systems theory, effectively allowing researchers to "linearize" the non-linear without local approximations[14,98].

Consider a dynamical network governed by the discrete-time non-linear evolution equation:

$$\mathbf{x}_{k+1} = \mathbf{f}(\mathbf{x}_k)$$

Analyzing this system directly in the state space $\mathbf{x} \in \mathbb{R}^n$ is analytically intractable due to the non-linearity of the map $\mathbf{f}$ . The Koopman operator, denoted by $\mathcal{K}$ , circumvents this by tracking the evolution of measurement functions—observables $g(\mathbf{x})$ —rather than the state vectors themselves. These observables may represent physical quantities, such as the instantaneous energy of a node or the phase synchronization order parameter.

The Koopman operator is defined as the composition of the observable with the dynamics, advancing the function forward in time:

$$\mathcal{K}g(\mathbf{x}_k) = g(\mathbf{f}(\mathbf{x}_k)) = g(\mathbf{x}_{k+1})$$

The profound utility of this definition lies in the fact that $\mathcal{K}$ is an infinite-dimensional linear operator, regardless of the non-linearity of $\mathbf{f}$. This linearity permits the application of spectral analysis—eigenvalues, eigenfunctions, and Singular Value Decomposition (SVD)—to systems that are classically chaotic or strongly non-linear.

Conceptually, this process constitutes a "lifting" of the dynamics. While trajectories in the original finite-dimensional state space may be entangled, knotted, or chaotic, lifting them into the infinite-dimensional space of observables untangles the geometry, rendering the flow as a linear evolution[3,4,99]. The central challenge of this framework is the infinite dimensionality of $\mathcal{K}$. Practical implementation necessitates the identification of a finite-dimensional invariant subspace spanned by the Koopman eigenfunctions, $\phi_j(\mathbf{x})$. These eigenfunctions define an intrinsic coordinate system where the complex dynamics decouple into simple modes of exponential growth, decay, or oscillation:

$$\phi_j(\mathbf{x}_{k+1}) = \lambda_j \phi_j(\mathbf{x}_k)$$

Consequently, the network dynamics can be decomposed into a superposition of these linear modes, revealing the coherent structures governing the system's evolution.

Dynamic Mode Decomposition (DMD) serves as the primary numerical algorithm for approximating the spectral properties of the Koopman operator from data[3,88,100]. Given a dataset of sequential network states, DMD constructs two data matrices: a snapshot matrix $\mathbf{X} = [\mathbf{x}_1, \dots, \mathbf{x}_{m-1}]$ and its time-shifted counterpart $\mathbf{X}' = [\mathbf{x}_2, \dots, \mathbf{x}_m]$. The algorithm seeks the optimal linear operator $\mathbf{A}$ that maps the system from current to future states, such that $\mathbf{X}' \approx \mathbf{A}\mathbf{X}$. The eigenvectors and eigenvalues of $\mathbf{A}$ approximate the Koopman modes and measuring the growth/decay rates of the dynamical features.

Standard DMD assumes that the state variables themselves ($\mathbf{x}$) form a sufficient basis for the observable space. Extended DMD enriches this space by employing a dictionary of non-linear basis functions (kernels), such as Hermite polynomials or radial basis functions[3,99]. This "kernel trick" allows EDMD to approximate the true Koopman operator with higher fidelity, capturing complex non-linearities that standard linear regression misses.

While standard DMD is agnostic to the underlying topology, GraphDMD explicitly integrates the network's adjacency structure into the decomposition. It imposes sparsity constraints derived from the graph Laplacian, ensuring that the learned modes respect the physical connectivity of the system[20,86,101]. This is particularly vital for applications like localized failure detection or cascading fault analysis.

Real-world networks—such as power grids monitored by Phasor Measurement Units (PMUs)—are frequently contaminated by non-Gaussian noise and outliers. Robust DMD leverages techniques from robust statistics and sparse optimization (akin to Robust PCA) to mathematically separate the low-rank coherent dynamics from sparse corruption. This prevents the model from overfitting to sensor errors or transient spikes[2,102,103].

A critical distinction exists between operator-theoretic methods and the geometric manifold learning techniques (e.g., Isomap, t-SNE). While geometric methods provide static mappings primarily useful for visualization and clustering, Koopman and DMD frameworks yield generative, predictive models[3,4]. They enable the forecasting of future states ( $\mathbf{x}_{t+T}$ ) and facilitate the design of linear control strategies for non-linear systems. However, the efficacy of these methods is contingent upon the "closure problem": if the selected dictionary of observables is insufficient to span an invariant subspace, the linear approximation will fail to capture the long-term dynamics.

## 4.4 Deep Geometric Learning

The convergence of Deep Learning and Dynamical Systems theory has catalyzed the development of highly scalable frameworks for network dimensionality reduction. In contrast to the spectral and geometric methods discussed previously—which predominantly rely on fixed matrix decompositions or iterative regression—these contemporary approaches leverage the differentiable programming paradigm. By parametrizing the reduction mapping as a deep neural network, these methods can learn complex, non-linear operators that generalize across different network topologies and dynamical regimes[4,5,10].

Variational Autoencoders (VAEs) represent a fundamental departure from deterministic dimensionality reduction. Rather than mapping an input $\mathbf{x}$ to a fixed point in the lower-dimensional space, VAEs employ stochastic variational inference to learn a probability distribution over the latent space, $q_\phi(\mathbf{z} \mid \mathbf{x})$. This probabilistic formulation introduces a regularization term (typically the Kullback-Leibler divergence from a Gaussian prior) that forces the latent space to be continuous and dense[3] . Consequently, VAEs exhibit inherent robustness to observation noise and enable generative modeling—the capacity to sample novel, synthetic network states that are statistically consistent with the training distribution.

Advanced VAE architectures, such as the $\beta$ -VAE, enforce stronger constraints to encourage "disentanglement." This objective seeks to learn orthogonal latent variables where each dimension of $\mathbf{z}$ corresponds to a distinct, interpretable factor of variation. In the context of network dynamics, a disentangled model might separate the "global synchronization order parameter" from "local community activity rates" into independent axes, rendering the latent space semantically interpretable.

Standard VAEs treat data points as independent and identically distributed (i.i.d.), ignoring temporal correlations. When applied to time-series, this often results in jagged, discontinuous trajectories where sequential states $t_1$ and $t_2$ are mapped to distant points in latent space due to noise. To mitigate this, Dynamical VAEs incorporate recurrence (e.g., LSTM or GRU cells) into the bottleneck or utilize loss functions that penalize high-order derivatives of $\mathbf{z}(t)$ [88]. This ensures the learned manifold is topologically smooth and that the system's evolution traces a continuous path.

Conventional deep learning architectures (such as CNNs or MLPs) possess an inductive bias towards grid-structured, Euclidean data, rendering them ill-suited for the irregular, non-Euclidean topology of complex networks. Graph Neural Networks (GNNs) address this limitation by explicitly operating on the graph structure[35]. GNNs update node embeddings via neural message passing, an iterative process where nodes aggregate feature information from their local topological neighbors. This mechanism grants GNNs two critical properties: permutation invariance (the ordering of nodes does not affect the output) and the ability to process graphs of varying sizes.

This hybrid architecture bridges discrete deep learning with continuous-time control theory. Rather than defining the network depth as a sequence of discrete layers, Neural ODEs parametrize the time derivative of the hidden state using a GNN[34,104]:

$$\frac{d\mathbf{h}(t)}{dt} = \mathrm{GNN}(\mathbf{h}(t), \mathbf{A}, \theta)$$

The output is computed by integrating this differential equation using an ODESolver. This formulation allows for the continuous-time modeling of dynamics, offering a robust solution for irregularly sampled time-series—a pervasive challenge in real-world sensor networks where data is often missing or asynchronous.

Representing the state-of-the-art in reduction, these models combine the probabilistic encoding of VAEs with the continuous dynamics of Neural ODEs. The framework first projects high-dimensional graph measurements into a compressed latent graph, evolves the dynamics within this latent space using a Neural ODE, and subsequently decodes the trajectory back to the full state space. This approach effectively captures both the topological complexity and the continuous temporal evolution in a highly compressed, computationally efficient form[4,105].

A decisive advantage of GNN-based methods over classical manifold learning (e.g., Isomap, LE) is the shift from transductive to inductive learning. Transductive methods (Isomap) learn a specific embedding for a fixed dataset; introducing a new node or changing the graph topology requires recomputing the entire embedding[1,2]. Inductive methods (GNNs) learn a generalized parametric function (an operator). Once trained, a GNN can be applied to unseen nodes, new subgraphs, or time-varying topologies without retraining. This capability is indispensable for real-time monitoring of evolving networks[101].

## 4.5 Scalability and Computational Complexity

The utility of any dimensionality reduction framework is inexorably bound to its computational scalability. While many spectral and geometric methods possess theoretical elegance, they often encounter insurmountable barriers when applied to modern networks where the number of nodes $N$ exceeds $10^5$. The "curse of dimensionality" here refers not only to the feature space but to the computational cost of resolving the topology of the sample space itself[1,2]. Table 4.1 provides a synthesis of the computational demands and key limitations for the full spectrum of discussed algorithms, ranging from classical spectral techniques to contemporary deep learning frameworks.

**Table 4.1: Comparative Analysis of Computational Complexity**

| **Algorithm** | **Nature** | **Complexity (Exact/Dense)** | **Complexity (Approx/Sparse)** | **Key Limitation for Dynamics** |
|---|---|---|---|---|
| **PCA[2,106]** | Linear | $O(N \cdot d^2 + d^3)$ | Randomized: $O(N \cdot d \cdot k)$ | Cannot unfold non-linear manifolds. |
| **MDS[1,107]** | Metric | $O(N^3)$ | Landmark: $O(M \cdot N)$ | No out-of-sample extension; slow. |
| **Laplacian Eigenmaps[8,11]** | Non-Linear (Spectral) | $O(N^3)$ | $O(k \cdot N)$ (Sparse Eig) | Collapses trajectories; no out-of-sample. |
| **Isomap[1,90]** | Non-Linear (Global) | $O(N^3)$ | Landmark: $O(M \cdot N \log N)$ | Sensitive to noise (short-circuiting). |
| **LLE** | Non-Linear (Local) | $O(N^2)$ | $O(k \cdot N \log N)$ | Distorts temporal/global aspect ratio. |
| **t-SNE[1,108]** | Non-Linear (Prob) | $O(N^2)$ | Barnes-Hut: $O(N\log N)$ | Shatters trajectories; no global metric. |
| **UMAP[109]** | Non-Linear (Topo) | $O(N^2)$ | $O(N)$ (NNDescent) | Can distort temporal density. |
| **DMD / Koopman[100]** | Linear Operator | $O(N\cdot m^2)$ | Randomized: $O(r\cdot N\log N)$ | Linear assumption; observable choice. |
| **SINDy[110]** | Sparse Regression | $O(N\cdot\Theta)$ | Parallelizable / STLS | Requires clean derivatives; basis selection. |
| **VAEs[10]** | Deep Probabilistic | $O(\text{Batch}\cdot W)$ | Scalable (GPU) | "Black box"; ignores topology. |
| **GNNs / Graph ODEs[35,101]** | Deep Graph | $O(E)$ (Edges) | Graph Sampling: $O(\text{Batch})$ | High training cost; over-smoothing. |

Classical kernel-based methods, including Isomap and Kernel PCA, rely fundamentally on global matrix operations. The precise determination of the embedding coordinates requires the spectral decomposition of an $N\times N$ dense matrix (e.g., the Gram matrix or the squared distance matrix)[1,107]. Standard eigendecomposition algorithms scale as $O(N^3)$, rendering exact solutions intractable for large datasets. Even when utilizing iterative sparse eigensolvers (such as Lanczos or Arnoldi iterations) to extract only the top-$d$ eigenvectors, the construction of the kernel matrix itself presents a quadratic bottleneck. For instance, Isomap requires computing all-pairs shortest paths; even with efficient implementations like Dijkstra (for sparse graphs) or Floyd-Warshall, the complexity remains between $O(N^2\log N)$ and $O(N^3)$.

To circumvent the $O(N^2)$ memory and processing limits, researchers employ low-rank matrix approximation strategies, most notably the Nyström extension and Landmark Isomap[2,111,112]. These techniques operate on the premise of redundancy: the geometry of the manifold can be sufficiently captured by a small, representative subset of $M \ll N$ "landmark" points. The algorithm computes the exact embedding for these landmarks (an $O(M^3)$ operation) and subsequently projects the remaining data points into the embedding space via interpolation kernels. This reduces the overall complexity to approximately $O(M^2N)$ or $O(MN)$, depending on the implementation, transforming the problem from cubic to linear with respect to the total dataset size.

The state-of-the-art in scalable reduction has shifted toward optimization-based frameworks that exploit sparsity. Algorithms like UMAP and large-scale Laplacian Eigenmaps construct $k$-Nearest Neighbor ($k$-NN) graphs rather than full distance matrices[8,11,86]. The naive construction of a $k$-NN graph is $O(N^2)$. However, modern Approximate Nearest Neighbors (ANN) algorithms, such as Hierarchical Navigable Small World (HNSW) graphs, facilitate neighbor search in $O(N \log N)$ or effectively $O(N)$ time for practical purposes[109]. Instead of solving a global eigenvalue problem, methods like UMAP and t-SNE optimize the embedding layout via Stochastic Gradient Descent (SGD). This allows the runtime to scale linearly with the number of edges in the $k$-NN graph, achieving near-linear scalability suitable for datasets with millions of nodes.

# 5 Discussion

The survey of methodologies presented in this review reveals a field at a critical epistemological juncture. The problem of dimensionality reduction in complex networks is no longer merely a challenge of computational compression; it has evolved into a fundamental debate regarding the nature of scientific explanation in the era of high-dimensional data. As we synthesize the landscapes of Structural Coarse-Graining, Analytical Reduction, and Data-Driven Inference, three dominant themes emerge: the tension between mechanistic derivation and phenomenological prediction, the inherent trade-offs between scalability and fidelity, and the necessary migration toward higher-order and inductive frameworks[1,2,113].

## 5.1 The Epistemological Schism

The most profound division in the current literature lies between Analytical (Equation-Based) and Statistical (Data-Driven) frameworks. This schism represents not merely a choice of computational tools, but a fundamental divergence in the epistemological goals of reduction (see Figure 5.1)[3,88].

The Analytical Imperative: Frameworks such as the Watanabe-Strogatz (WS) ansatz or the Gao-Barzel-Barabási (GBB) method are driven by the ambition to derive a "macroscopic theory" directly from microscopic laws[6,9]. By seeking a transformation that explicitly maps the full state space to a reduced vector field, these methods preserve the causal link between network topology and dynamical function. This preservation allows for the rigorous derivation of stability boundaries and resilience patterns, enabling researchers to predict critical transitions without resorting to brute-force simulation. However, the practical application of these derivations often hinges on the successful resolution of the "closure problem"—the mathematical necessity of truncating the infinite hierarchy of moment equations, where the evolution of low-order moments invariably depends on higher-order terms[6,9].

The Data-Driven Turn: Conversely, the field has increasingly pivoted toward data-driven techniques that infer low-dimensional representations directly from observational data, bypassing the requirement for explicit mechanistic models. Methods like Variational Autoencoders (VAEs), t-SNE, and Koopman operators excel at capturing the variance and topological structure of the data[10,65,93,97]. However, these approaches are often fundamentally phenomenological; they generate compressed representations of what occurred—the trajectory—without elucidating why it occurred—the mechanism[1,5,114]. Furthermore, as highlighted in Section 4.1, standard linear manifold learning techniques like PCA can catastrophically overestimate intrinsic dimensions if the manifold is non-linear, or erroneously filter out critical transient signals (such as bifurcation precursors) as noise during the variance-maximization process.

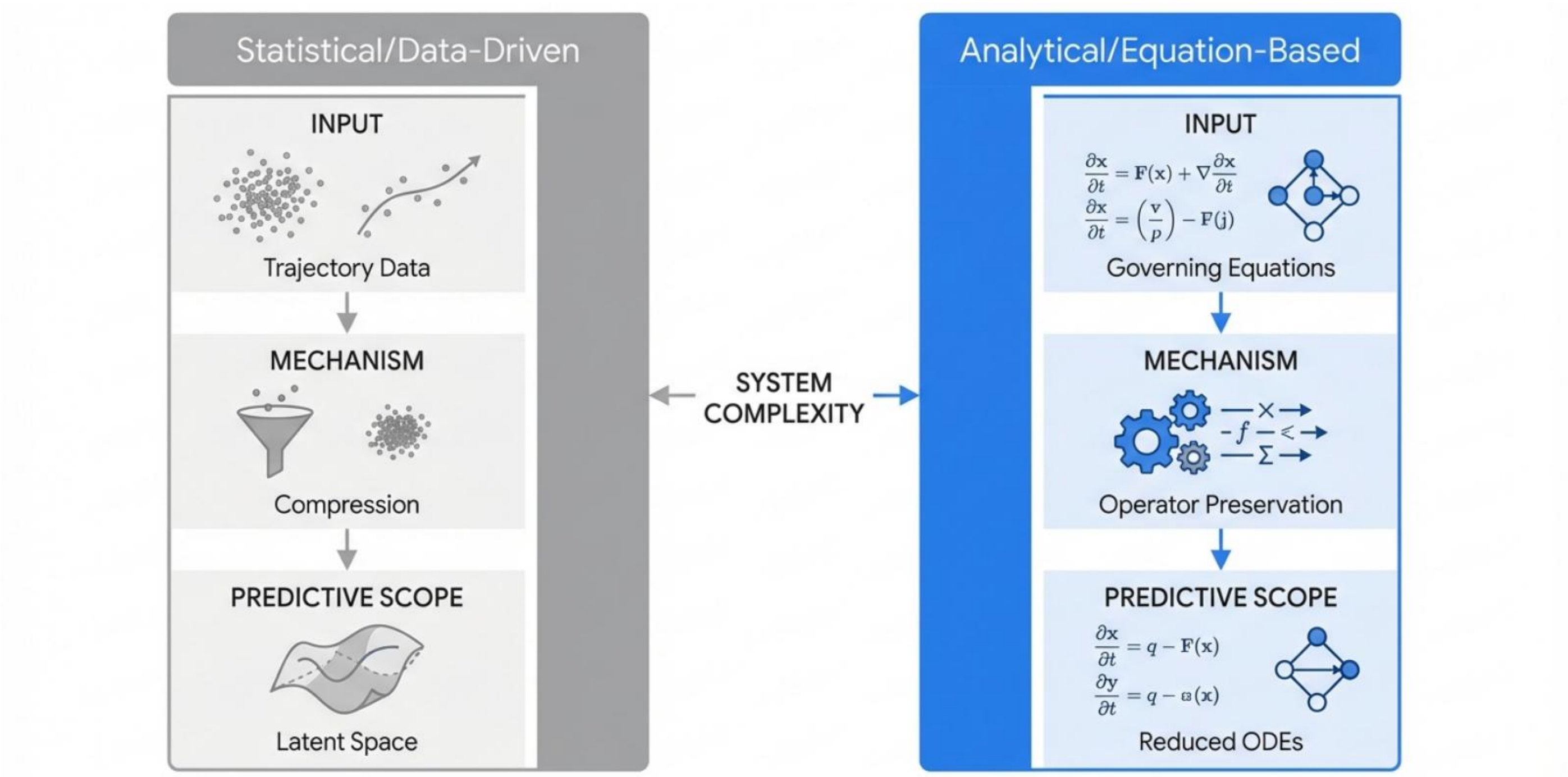

**Figure 5.1: Taxonomy of dimensionality reduction in dynamical systems.** A comparative schematic contrasting statistical (data-driven) and analytical (equation-based) frameworks. On the left, data-driven methods utilize trajectory data inputs to derive compressed latent spaces. On the right, analytical methods operate directly on the governing equations via operator preservation to generate reduced Ordinary Differential Equations (ODEs), thereby retaining the explicit link between structure and dynamics.

## 5.2 The Pareto Frontier of Fidelity and Scalability

The comparative analysis reveals a rigorous "No Free Lunch" theorem governing the selection of reduction algorithms. The field is defined by a fundamental trade-off between computational tractability and physical fidelity. This trade-off establishes a Pareto frontier of reduction methodologies, where gains in scalability are almost invariably purchased at the expense of theoretical exactness[1,2].

The Trap of Exactness: Symmetry-based reductions, such as the Watanabe-Strogatz ansatz, occupy the extreme of exactness[9]. They rely on the rigorous mathematical properties of automorphism groups to preserve full transient dynamics and soliton solutions without error. However, this precision comes with significant fragility. Their reliance on specific algebraic structures—such as Möbius invariance or strict topological symmetry—restricts their validity to highly idealized systems[45]. They are inherently brittle; a single misplaced edge or a deviation in natural frequencies can destroy the automorphism group, invalidating the reduction entirely[115].

The Cost of Spectral Rigor: Spectral Coarse Graining (SCG) provides a robust alternative to exact methods by accepting approximation errors in exchange for resilience to disorder[8,16]. By targeting isospectral fidelity, SCG ensures that global relaxation timescales and synchronization thresholds remain invariant[116]. However, this theoretical rigor imposes a severe computational penalty. The requisite eigen-decomposition introduces an $O(N^3)$ bottleneck for dense matrices[11,86]. Even when utilizing sparse iterative solvers to reduce complexity to $O(k \cdot E)$, the computational cost remains prohibitive for massive networks exceeding $10^7$ nodes, particularly in modular graphs where spectral gaps are small[117].

The Scalability of Local Approximation: To breach the scalability barrier, frameworks like Iterative Structural Coarse Graining (ISCG) abandon global spectral analysis in favor of local motif identification[16]. By exploiting the modular architecture of complex networks and contracting local cliques or $k$-plexes, ISCG achieves near-linear scalability ($O(N)$)[118]. This computational efficiency renders ISCG the only viable option for massive epidemiological simulations where global matrix operations are intractable. However, this "local" approach sacrifices global spectral precision, trading the exact preservation of diffusion modes for the ability to model non-linear contagion on the scale of millions of nodes[69,119].

## 5.3 Bridging the Gap: Hybrid and Inductive Architectures

Looking forward, the next generation of reduction frameworks must address two critical limitations that currently bifurcate the field: the rigidity of transductive embeddings and the opacity of black-box

inference[1,2,114].

From Transductive to Inductive Learning: A fundamental limitation of classical manifold learning techniques—such as Isomap, MDS, and Laplacian Eigenmaps—is their transductive nature[90,92,120]. These algorithms optimize an embedding specific to a fixed dataset; they do not learn a mapping function, but rather a set of coordinates. Consequently, they lack a direct out-of-sample extension: if the network topology evolves or new nodes are introduced, the entire optimization must be recomputed from scratch[1]. The paradigm shift toward Graph Neural Networks (GNNs) represents a critical advancement because these architectures are inherently inductive. Instead of optimizing node coordinates directly, GNNs learn a generalized parametric operator (a set of message-passing filters). Once trained, this operator can be applied to unseen nodes, new subgraphs, or time-varying topologies without retraining[101]. This capability is indispensable for the real-time monitoring of evolving networks, allowing for the continuous projection of dynamic states[10].

Physics-Informed Hybridization: The most promising frontier lies in the synthesis of equation-based priors with deep learning, often categorized under the banner of Scientific Machine Learning[5,13]. Methodologies such as Neural ODEs and G-LED (Generative Learning of Effective Dynamics) exemplify this approach[104]. Rather than discarding physical knowledge in favor of pure pattern recognition, these hybrid models use neural networks to augment analytical derivations. For instance, Neural ODEs parametrize the time-derivative of the state vector with a neural network, allowing the model to learn complex, continuous-time dynamics that fit irregularly sampled data[105]. Similarly, frameworks like G-LED utilize data-driven inference to "learn" the effective dynamics that correct for the truncation errors inherent in mean-field analytical approximations[3,121]. By integrating the structural inductive bias of differential equations with the adaptive accuracy of deep learning, these architectures solve the "closure problem" computationally, combining mechanistic interpretability with high-fidelity prediction[5,13].

## 5.4 Beyond the Dyadic and Static Paradigm

Finally, this review underscores a critical imperative: the methodologies of dimensionality reduction must co-evolve with the increasing sophistication of network ontology itself. Standard graph-theoretic reductions are rapidly reaching their limits as the modeling of complex systems transcends simple static graphs[1,2].

Higher-Order Interactions: The scientific consensus is pivoting away from the reductionist reliance on purely dyadic (pairwise) interaction models. In domains ranging from neural coding to social collaboration, the fundamental unit of interaction is polyadic—best represented mathematically by Simplicial Complexes and Hypergraphs[122,123]. Consequently, reduction techniques must transcend standard vertex-based aggregation.

The development of Higher-Order Laplacian Renormalization is essential to capture the multi-scale topological invariants (such as Betti numbers and simplicial flux) that remain invisible to traditional metrics[15]. As noted in the literature, a network may exhibit scale-free behavior in its node degrees while displaying entirely distinct, non-trivial scaling laws in its higher-order connectivity—features that only higher-order spectral analysis can reveal[124,125].

Geometric Embeddings: Simultaneously, the conceptualization of complex networks as discrete discretizations of underlying continuous manifolds offers a powerful theoretical direction[52,126]. Specifically, the Geometric Renormalization Group (GRG) posits that the latent geometry of complex networks is often characterized by hyperbolic curvature. This framework provides a theoretical unification for reduction processes. It suggests that dimensionality reduction is physically equivalent to "zooming out" or radial rescaling within this latent hyperbolic space. Remarkably, this geometric renormalization flow naturally generates the ubiquitous structural hallmarks of real-world networks, such as high clustering coefficients and small-world path lengths, framing them not as random artifacts but as inevitable consequences of the underlying hyperbolic geometry[127-129].

# 6 Conclusion

The scientific endeavor to map the function of complex systems is fundamentally an exercise in compression. As we have detailed, the dimensionality reduction of network dynamics is not merely a computational convenience; it is an epistemological necessity for extracting the "effective laws" that emerge from high-dimensional chaos. This review has synthesized the three dominant methodological lineages—structural, analytical, and data-driven reduction—revealing a landscape defined by the inescapable trade-off between physical fidelity and computational scalability.

We first examined Structural Coarse-Graining, which seeks to reduce complexity by physically contracting the network topology. While methods like Spectral Coarse Graining (SCG) offer rigorous preservation of linear diffusive modes and synchronization thresholds, they are often hindered by the cubic complexity of eigen-decomposition. Conversely, motif-based approaches like Iterative Structural Coarse Graining (ISCG) have emerged as scalable alternatives for epidemiological modeling, effectively trading spectral precision for the ability to handle massive, heterogeneous datasets.

Parallel to topological contraction, Analytical Equation-Based Reduction aims to compress the state space itself. The canonical reductions of Watanabe-Strogatz and Ott-Antonsen demonstrated that infinite-dimensional oscillator systems could be exactly described by low-dimensional ODEs, provided specific symmetries hold. To address the heterogeneity of real-world networks, effective operator frameworks (GBB) and statistical moment closures (HMF, PA) were developed. These methods provide profound mechanistic insights—linking topology directly to resilience—but are mathematically constrained by the "closure problem," where the truncation of higher-order correlations limits their validity near critical phase transitions.

Finally, the review addressed the paradigmatic shift toward Data-Driven Reduction. Techniques ranging from linear PCA to non-linear Manifold Learning (Isomap, t-SNE) and Deep Geometric Learning (GNNs, Neural ODEs) have bypassed the need for explicit governing equations. While these frameworks offer inductive scalability and can model systems with unknown laws, they often remain phenomenological "black boxes," lacking the causal interpretability necessary for robust control strategies.

Looking forward, the frontier of this field lies in the reconciliation of these divergent paths. The "schism" between data-driven and equation-based methods is beginning to close through the emergence of hybrid frameworks. Innovations such as Generative Learning of Effective Dynamics (G-LED) and robust operator-theoretic methods (Robust DMD) illustrate a future where deep learning is used to learn the "closure" terms that analytical theory cannot derive, while physical priors ensure the resulting models remain consistent and interpretable. Ultimately, the goal is to move beyond simple compression to true macroscopic discovery: enabling us to see not just the data, but the dynamical skeleton of complexity itself.

# Acknowledgements

This paper is supported by Zhejiang Provincial Philosophy and Social Sciences Planning Project (Grant No. 24NDJC175YB). C.T. acknowledges support from the National Natural Science Foundation of China (Grant No. 72571247) and Scientific Research Project of Zhejiang Provincial Bureau of Statistics (Grant No. 25TJZZ18).

# Conflicts of Interest

The authors declare no conflicts of interest.